\documentclass[11pt]{article} 
\usepackage{graphicx,amssymb,mathrsfs,array,multirow}  
\begin{document}  
 
\vskip 30pt

\begin{center}  
{\Large{\bf Three-Higgs-doublet model under $A4$ symmetry implies
alignment}}\\
\vspace*{1cm}  
\renewcommand{\thefootnote}{\fnsymbol{footnote}}  
{ {\sf Soumita Pramanick$^{1,2}$\footnote{email: soumita509@gmail.com}},
{\sf Amitava Raychaudhuri$^1$\footnote{email: palitprof@gmail.com}} 
} \\  
\vspace{10pt}  
{\small  {\em $^1$Department of Physics, University of Calcutta,  
92 Acharya Prafulla Chandra Road, Kolkata 700009, India\\
$^2$Harish-Chandra Research Institute, Chhatnag Road, Jhunsi,
Allahabad 211019, India \\}}
\normalsize

\end{center}  

\begin{abstract} 

{\it A model with three scalar doublets can be conveniently
accommodated within an $A4$ symmetric framework. The $A4$
symmetry permits only a restricted form for the scalar potential.
We show that for the global minima of this potential  alignment
follows as a natural consequence. We also verify that in every
case positivity and unitarity constraints are satisfactorily met.
}

\vskip 5pt \noindent  
\end{abstract}  

\renewcommand{\thesection}{\Roman{section}} 
\setcounter{footnote}{0} 
\renewcommand{\thefootnote}{\arabic{footnote}} 
\noindent

\section{Introduction}

The discovery at the LHC in 2012 of a spin-0 particle of mass
around 125 GeV \cite{Higgs} with properties closely matching that of the
Standard Model (SM) Higgs boson is a major vindication of our
understanding of the elementary particle properties. This colour
neutral particle arises from a multiplet which transforms under
the $SU(2)_L$ symmetry as a doublet. This is precisely what is
required to generate masses for quarks and leptons and the gauge
bosons while keeping $M_W/M_Z = \cos\theta_W$ in agreement
with observations.

In spite of this success, the scalar sector of the Standard Model
still retains several directions which merit investigation. A
much discussed issue is that of naturalness, namely, there is no
obvious reason for the protection of a light Higgs scalar mass.
Different directions of addressing this impasse such as
supersymmetry, compositeness, extra dimensions, clockwork, etc. have been
under examination and experimental tests for these alternatives
are being  pursued with vigour.\footnote{There are other
shortcomings of the Standard Model such as  massless neutrinos
and the lack of a dark matter candidate. Our focus in this work
will be restricted to the scalar sector.}

Besides naturalness there is also the question of minimality of
the scalar field content. Is there only one scalar doublet as
postulated in the Standard Model?  Even though one scalar doublet
serves most purposes rather well could there be in addition
further scalar multiplets transforming under $SU(2)_L$ either as
doublets or as other representations? The simplest extension
could be the addition of $SU(2)_L$ singlet scalars
\cite{anindya}.  Alternatively, seesaw models of neutrino mass of
the Type-II variety rely on the introduction of an $SU(2)_L$
triplet scalar multiplet
\cite{T2seesaw}.  But by far the most attention has been devoted
to multi-doublet models among which justifiably the simplest
two-Higgs-doublet model has been  covered most extensively
\cite{2hdmrev}. A specially important sub-class of these are the
supersymmetric models which necessitate two $SU(2)_L$ doublets
with rather specific couplings. Models with $n$-Higgs doublets
with $n > 2$ have also been under study \cite{multi, haber2}.

In this work we consider a model in which there are three scalar
doublets which transform as a triplet under the discrete symmetry
$A4$. Models with three Higgs doublets have been of interest in
their own right and have been examined from various angles
\cite{3hdmother}. The scalar spectrum of such models with
possible discrete and continuous symmetries have been
investigated in \cite{King1} while the potential minima and
CP-violation options have been examined in \cite{King2, King3}.
For a three-Higgs-doublet model with $S3$ symmetry novel scalar
decays \cite{S3gb},  the spectrum of the scalar sector and its
consequences \cite{S3dd}, and the high energy behaviour of the
potential \cite{S3nc} have been explored.  $A4$ as a flavour symmetry for
lepton and quark masses was first considered in \cite{A4mr} who
introduced three scalar doublets transforming as a triplet of
the $A4$ group and wrote down the most general potential
consistent with the symmetry. They showed that a choice
of the  
symmetry breaking where the vacuum expectation value (vev) for
the three doublets were  equal led to a lepton mass model with
attractive features. Closer to the
spirit of this work, the authors in \cite{A4Toorop} also consider
the same 
model with three scalar doublets transforming as an $A4$ triplet. 
They extract the scalar mass spectra for different vacuum
expectation value patterns, to which our
calculations agree, and examine their implications on gauge boson
decays and on oblique corrections. Our
primary focus in this work is different; it is to establish the
``alignment" feature as discussed below.  Another work with the
same particle content but  with soft symmetry breaking terms has
been the subject of \cite{A4Higgs}. A model with several
$A4$-triplet scalars and additional discrete symmetries has also
been studied \cite{A4disc}.

We consider a model with one $A4$-triplet consisting of
$SU(2)_L$-doublet scalars. We do not allow any soft $A4$ breaking
terms.  Demanding that the scalar potential respects $A4$
symmetry imposes restrictions on the allowed terms and relates
them.  We find that for all global minima of the potential these
relations automatically imply vacuum alignment without any
fine-tuning whatsoever. In every
case in the mass eigenstate basis of the scalar fields, the
so-called `Higgs basis', the vacuum expectation value  is
restricted to only one of the three multiplets \cite{haber}. This
multiplet has a massive neutral scalar, i.e., the SM Higgs boson
analogue, and a massless neutral and a massless charged scalar,
the Goldstone modes\footnote{Fermion masses, which are beyond the
scope of this paper,  also arise from their coupling to this
multiplet.}.  The other mass eigenstate scalars, all of
non-zero mass, are
superpositions of the remaining two scalar $SU(2)_L$ doublets,
with their exact composition varying case by case. We discuss the
consequences on the model from requiring positivity of the
potential and also demanding that tree-level $s$-wave unitarity be
satisfied.

We stress here that this is at best only a toy model.
Since the magnitude of the effective vev, $v$, is controlled by
the gauge boson masses and it is the only mass parameter in the
model, all scalars end up with either vanishing mass (the
Goldstone states) or have mass ${\cal O}(v)$. Realistic models
which incorporate quark and lepton masses usually have a richer
scalar sector \cite{A4mr, A4af, OurA4}.

In the next section we briefly review the $A4$ symmetry group. In
the following section we write down the $A4$-symmetric scalar potential
of the three-doublet model. The physics consequences of this
model are presented in the two next sections where we discuss the
scalar masses and alignment and the bounds
arising from positivity and unitarity. We end with our
conclusions and discussions.

\section{The $A4$ group}

The discrete group $A4$ comprises of twelve elements
corresponding to the  even permutations of four objects.  Two
basic permutations $S$ and $T$ which satisfy
$S^2=T^3=(ST)^3=\mathbb{I}$ and their nontrivial products
generate $A4$. The inequivalent irreducible representations are
four in number; one of 3 dimension and three of 1 dimension which
we denote by $1, 1'$ and $1''$. The latter are singlets under $S$
and transform under $T$ as 1, $\omega$, and $\omega^2$
respectively, where $\omega$ is a complex cube root of unity.
The one-dimensional representations satisfy
\begin{equation}
1' \times 1'' = 1 \;\;. 
\label{A41'x1''}
\end{equation}

For the remaining representation of dimension 3 one has
\begin{equation}
S=\pmatrix{1 & 0 & 0 \cr 0 & -1 & 0 \cr 0 & 0 & -1}\ \ \ \ {\rm and} \ \ \ \
T=
\pmatrix{0 & 1 & 0 \cr
0 & 0 & 1 \cr
1 & 0 & 0} \;\;.
\label{ST3}
\end{equation}
As is seen from the above, in this basis the generator $S$ is
diagonal. We will use this basis. It is noteworthy that in the
literature a basis in which the generator $T$ is diagonal (with
eigenvalues $1, \omega, \omega^2$) has also been used. The two
bases are related by a unitary transformation by:
\begin{equation}
U_3 = \frac{1}{\sqrt{3}} \pmatrix{1 & 1 & 1 \cr 1  &
\omega^2 & \omega \cr 1 & \omega &  \omega^2}  \;.
\label{u3}
\end{equation} 
This matrix will  reappear in our discussions later.

For the 3-dimensional representation the product rule is
\begin{equation}
3\otimes3=1 \oplus 1' \oplus 1'' \oplus 3 \oplus 3 \;\;.
\label{A43x3}
\end{equation}
The triplets $3_c, 3_d$ arising from the product
of two triplets $3_a \equiv {a_i}$ and $3_b \equiv {b_i}$, where 
$i=1,2,3$, can be represented as
\begin{equation}
c_i = (a_2 b_3, a_3 b_1, a_1 b_2) \;\;{\rm and} \;\;  
d_i = (a_3 b_2, a_1 b_3, a_2 b_1) \;\;.
\label{3x3to3}
\end{equation}
In the same notation the other
representations in the $3_a\otimes3_b$ product are: 
\begin{equation}
1 = a_1b_1+a_2b_2+a_3b_3  \;\;,\;\;  
1' = a_1b_1+\omega^2a_2b_2+\omega a_3b_3 \;\;,  \;\;
1'' = a_1b_1+\omega a_2b_2+\omega^2a_3b_3  \;\;. 
\label{3x3to1}
\end{equation}
More details of the 
$A4$ group can be found in \cite{A4mr, A4af}.

Models of quarks and leptons based on $A4$ as a flavour symmetry
group have been widely examined. Typical examples of such
applications for the issue of neutrino masses can be found in
\cite{A4mr, A4af, OurA4, neutrino} and those for quark masses in
\cite{quarks1, quarks2}. 

\section{The $A4$-symmetric scalar sector}

We consider here a model where there is one scalar multiplet
which transforms as an $A4$ triplet. The components of this
muliplet are colour neutral and under the electroweak symmetry
transform as three $SU(2)_L$ doublets each with hypercharge $Y =
1$. We represent this collection of scalars as:
\begin{equation}
\Phi \equiv \pmatrix{\Phi_1 \cr \Phi_2
\cr \Phi_3} \equiv 
\pmatrix{\phi_1^+ & \phi_1^0 \cr \phi_2^+ & \phi_2^0 
\cr \phi_3^+ & \phi_3^0 } \;\;,
\label{multipletp}
\end{equation}
where the $SU(2)_L$ symmetry acts horizontally while the $A4$
transformations do so vertically. We decompose the neutral fields
into the scalar and pseudoscalar components: $\phi_i^0 = \phi_i +
i \chi_i$.

Our objective is to explicitly show that alignment holds for
the vev which have been identified as the global minima of the
potential. This implies \cite{haber} that there exists a unitary
transformation  $U$ such that if
\begin{equation}
U \pmatrix{\Phi_1 \cr \Phi_2
\cr \Phi_3} = \Psi \equiv \pmatrix{\Psi_1 \cr \Psi_2
\cr \Psi_3} \equiv 
\pmatrix{\psi_1^+ & \psi_1^0 \cr \psi_2^+ & \psi_2^0 
\cr \psi_3^+ & \psi_3^0 } \;\;,
\label{Higgsbasis}
\end{equation}
then in the $\Psi$ basis the vev is restricted to only one
component, $\langle \psi_i^0 \rangle \neq 0$ and $\langle \psi_j^0
\rangle = 0$ for $j \neq i$. At the same time, the members of
$\Psi_i$, namely, $\psi_i^+$ and $\psi_i^0 \equiv \eta_i^0 + i
\xi_i^0$, are mass eigenstates with a massive neutral state and
one massless neutral state along with a massless charged state.
The other mass eigenstates are superpositions of the remaining
states $\Psi_j ~~(j \neq i)$. For most purposes $\Psi_i$ mimics
the Standard Model Higgs scalar doublet.

\subsection{The scalar potential}
We will express the potential in terms of the components
$\Phi_i$, each of which is an $SU(2)_L$ doublet scalar multiplet.
$A4$-symmetry obviously implies a unique quadratic term, i.e.,
the same mass term for all components. No cubic terms are
permitted by the electroweak symmetry. Turning now to the
quartics it is useful to consider first the product of the
triplet with itself and then the product of two such
combinations. According to Eq. (\ref{A43x3}) the product of two
$A4$ triplets can give rise to two triplets ($3_c$ and $3_d$ in
Eq. (\ref{3x3to3})) besides a $1$, a $1'$, and a $1''$. Out of
these, in the quartic term the two singlets together form a
singlet as do the $1'$ with the $1''$ -- see  Eq.  (\ref{A41'x1''}).
Two triplets can form a singlet but out of the four possibilities
arising from $3_c$ and $3_d$ only two are independent. These are
all the quartic terms allowed by the gauge and discrete symmetry.
The potential in terms of components is then \cite{A4mr}:
\begin{eqnarray}
V(\Phi_i) &=& m^2 \left(\sum_{i=1}^3 \Phi_i^\dagger \Phi_i\right)  + 
\frac{\lambda_1}{2} \left(\sum_{i=1}^3 \Phi_i^\dagger \Phi_i\right)^2  
\nonumber \\ &+ &  
\frac{\lambda_2}{2} \left(\Phi_1^\dagger \Phi_1 +
\omega^2 \Phi_2^\dagger \Phi_2 + \omega \Phi_3^\dagger \Phi_3 \right)
\left(\Phi_1^\dagger \Phi_1 +
\omega \Phi_2^\dagger \Phi_2 + \omega^2 \Phi_3^\dagger \Phi_3 \right)  
\nonumber \\ &+ &  
 \frac{\lambda_3}{2} \left[ \left(\Phi_1^\dagger \Phi_2\right)
\left(\Phi_2^\dagger \Phi_1\right) +
\left(\Phi_2^\dagger \Phi_3\right) \left(\Phi_3^\dagger \Phi_2\right) +
\left(\Phi_3^\dagger \Phi_1\right)\left(\Phi_1^\dagger \Phi_3\right) \right]
\nonumber \\ &+ &  
 \lambda_4 \left[ \left(\Phi_1^\dagger \Phi_2\right)^2 + 
\left(\Phi_2^\dagger \Phi_1\right)^2 +
\left(\Phi_2^\dagger \Phi_3\right)^2 +  \left(\Phi_3^\dagger \Phi_2\right)^2 +
\left(\Phi_3^\dagger \Phi_1\right)^2 + \left(\Phi_1^\dagger \Phi_3\right)^2
 \right]
\;\;.
\label{V1}
\end{eqnarray}
We take all $\lambda_i ~(i =1,\ldots 4)$ to be real. In general,
only $\lambda_4$ can be complex. We comment, in passing, on the
impact of this option.

We can rewrite the second term by using the property $1+ \omega +
\omega^2 = 0$ to get:
\begin{eqnarray}
V(\Phi_i) &=& m^2 \left(\sum_{i=1}^3 \Phi_i^\dagger \Phi_i\right)  + 
\frac{\lambda_1 + \lambda_2}{2} \left(\sum_{i=1}^3 \Phi_i^\dagger
\Phi_i\right)^2
\nonumber \\ & - &  
\frac{3 \lambda_2}{2}  \left[ \left(\Phi_1^\dagger \Phi_1\right)
\left(\Phi_2^\dagger \Phi_2\right) +
\left(\Phi_2^\dagger \Phi_2\right) \left(\Phi_3^\dagger \Phi_3\right) +
\left(\Phi_3^\dagger \Phi_3\right)\left(\Phi_1^\dagger \Phi_1\right) \right]
\nonumber \\ &+ &  
 \frac{\lambda_3}{2} \left[ \left(\Phi_1^\dagger \Phi_2\right)
\left(\Phi_2^\dagger \Phi_1\right) +
\left(\Phi_2^\dagger \Phi_3\right) \left(\Phi_3^\dagger \Phi_2\right) +
\left(\Phi_3^\dagger \Phi_1\right)\left(\Phi_1^\dagger \Phi_3\right) \right]
\nonumber \\ &+ &  
 \lambda_4 \left[ \left(\Phi_1^\dagger \Phi_2\right)^2 + 
\left(\Phi_2^\dagger \Phi_1\right)^2 +
\left(\Phi_2^\dagger \Phi_3\right)^2 +  \left(\Phi_3^\dagger \Phi_2\right)^2 +
\left(\Phi_3^\dagger \Phi_1\right)^2 + \left(\Phi_1^\dagger \Phi_3\right)^2
 \right]
\;\;.
\label{V2}
\end{eqnarray}
We will be using this form in the subsequent calculations.

\section{The four alternative global minima}\label{vevcases}

For spontaneous symmetry breaking the neutral scalar fields in
$\Phi$ develop vacuum expectation values. 
The following alternatives have been shown to be the only possible
global minima of the potential \cite{gmin1, gmin2} and have commonly
appeared in the literature\footnote{If $\lambda_4$ is complex
then a more general form $\langle \Phi \rangle_2 =
\frac{v}{2} \pmatrix{0 & 1 \cr 0 & e^{i\alpha} \cr 0 & 0}$
is possible, where $\sin 2\alpha \propto Im(\lambda_4)$ as
discussed in the Appendix.}:
\begin{equation}
\langle \Phi \rangle_1 = \frac{v}{\sqrt{2}} \pmatrix{0 & 1 \cr 0
& 0 \cr 0 & 0} \;,\;
\langle \Phi \rangle_2 = \frac{v}{2} \pmatrix{0 & 1 \cr 0
& 1 \cr 0 & 0} \;,\;
\langle \Phi \rangle_3 = \frac{v}{\sqrt{6}} \pmatrix{0 & 1 \cr 0
& 1 \cr 0 & 1} \;,\;
\langle \Phi \rangle_4 = \frac{v}{\sqrt{6}} \pmatrix{0 & 1 \cr 0
& \omega \cr 0 & \omega^2} \;.
\label{vevs}
\end{equation}
Here $v = v_{SM} \sim 246$ GeV. We examine each of these options
in turn. We determine the condition under which any particular
minimum arises from Eq.  (\ref{V2}) and then work out the mass
matrices of the physical scalars that emerge. For this we use the
following notation: 
\begin{equation}
{\cal L}_{mass} = \frac{1}{2}(\chi_1 ~\chi_2 ~\chi_3) ~M^2_{\chi_i \chi_j}
\pmatrix{\chi_1 \cr \chi_2 \cr \chi_3}
+ \frac{1}{2} (\phi_1 ~\phi_2 ~\phi_3) ~M^2_{\phi_i \phi_j}
\pmatrix{\phi_1 \cr \phi_2 \cr \phi_3} 
 +  (\phi^-_1 ~\phi^-_2 ~\phi^-_3) ~M^2_{\phi^\mp_i \phi^\pm_j}
\pmatrix{\phi^+_1 \cr \phi^+_2 \cr \phi^+_3}  
\label{mdef}
\end{equation}

In every case we verify that alignment is a consequence.

\subsection{Case 1: $\langle\phi_i^0\rangle = \frac{v}{\sqrt{2}} (1,0,0)$} 

We begin with the case where $\langle\phi_1^0\rangle = \frac{v}{\sqrt{2}}$
and $\langle\phi_2^0\rangle = \langle\phi_3^0\rangle = 0$, i.e.,
\begin{equation}
\langle \Phi \rangle_1 = \frac{v}{\sqrt{2}} \pmatrix{0 & 1 \cr 0
& 0 \cr 0 & 0}  \;.
\label{vev1}
\end{equation}
In this case alignment will be true if the components of
$\Phi_1$ be mass eigenstates, of which the charged and a neutral
scalar become Goldstone modes of zero mass. We show that this is
indeed the case.

From Eq. (\ref{V2}) we find that the minimisation condition that
must be satisfied for the vev in Eq. (\ref{vev1}) is:
\begin{equation}
m^2 + \frac{v^2}{2} \left[\lambda_1  + \lambda_2\right] = 0  \;.
\label{min1}
\end{equation}

Using this condtion and the full scalar potential in Eq.
(\ref{V2}) we can find the mass matrices for the charged scalars
($\phi_i^\pm$) and the neutral scalars ($\phi_i$) and
pseudoscalars ($\chi_i$).  The $ij$-th off-diagonal entry of any
mass matrix depends on the combination $v_i v_j$  and since in
this case $v_2 = v_3 = 0$ the mass matrices are all diagonal.

For the charged scalar sector the mass-squared matrix is:
\begin{equation}
 M^2_{\phi^\mp_i \phi^\pm_j} = diag(0 \;,\; r_+ \;,\; r_+) \;\; {\rm
where}  \;\;  r_+ =
\frac{v^2}{4} ~(-3  \lambda_2) \;\;.
\label{chgd1}
\end{equation}
The Goldstone state $\phi_1^\pm$ becomes the longitudinal
mode of the charged gauge boson. The mass-squareds of the two
remaining degenerate states will be positive if $\lambda_2
< 0$. We show in the next section that such a choice is consistent
with the positivity of the potential and in agreement with
unitarity bounds.  Since the couplings
$|\lambda_i| \leq {\cal O}(16 \pi)$ from perturbativity, the
massive charged scalars can be in the 100 GeV to a TeV
range.

The vev and the $\lambda_i$ being real the neutral scalar
($\phi_i$) and pseudoscalar ($\chi_i$) sectors remain
independent. We get for the neutral pseudoscalars:
\begin{equation} M^2_{\chi_i \chi_j} = diag(0 \;,\; p \;,\; p) \;\; {\rm
where}  \;\; p = \frac{v^2}{4} \left(-3 \lambda_2 + \lambda_3 - 4
\lambda_4\right) \;\;.
\label{imag1}
\end{equation}
We can readily identify the zero mass Goldstone mode, $\chi_1$,
while $\chi_{2,3}$ are massive degenerate states. We show in the following
section that positivity and unitarity constraints do allow
positive mass-squareds for these scalars.

Finally, for the neutral real scalars we have:
\begin{equation}
 M^2_{\phi_i \phi_j} = diag(q \;,\; r_0 \;,\; r_0) \;\; {\rm where}
\;\;  q =
v^2 ~(\lambda_1 + \lambda_2) \;,\;  r_0  = \frac{v^2}{4} \left(-3
\lambda_2 + \lambda_3 + 4 \lambda_4\right) \;\;.
\label{real1}
\end{equation}
Positivity of the scalar potential requires $(\lambda_1 +
\lambda_2)$ to be positive. So, $\phi_1$ has a positive
mass-squared. Further, $r_0 = m^2_{\phi_2, \phi_3}$ is also
positive.  In other words, alignment is manifest and the unitary
transformation in Eq. (\ref{Higgsbasis})  for this case is
the unit matrix. The defining basis is also the Higgs basis.

If we had taken $\lambda_4$ to be complex, then the charged
sector would be unaffected as would be $\phi_1$ and $\chi_1$. The
other mass
eigenstates would be orthogonal superpositions of $\phi_2$ with
$\chi_2$ and $\phi_3$ with $\chi_3$, the mixing angle being
proportional to $Im(\lambda_4)$.

\subsection{Case 2: $\langle\phi_i^0\rangle = \frac{v}{2}
(1,1,0)$}\label{case2}

This is the global minimum for which  $\langle\phi_1^0\rangle = \langle\phi_2^0\rangle =
\frac{v}{2}$ and $\langle\phi_3^0\rangle = 0$, i.e.,
\begin{equation}
\langle \Phi \rangle_2 = \frac{v}{2} \pmatrix{0 & 1 \cr 0
& 1 \cr 0 & 0}  \;.
\label{vev2}
\end{equation}

One can get this minimum if the potential satisfies:
\begin{equation}
m^2 + \frac{v^2}{4} \left[ \lambda_1 + \frac{1}{4} \lambda_2
+ \frac{1}{4} \lambda_3 + \lambda_4 \right] = 0\;.
\label{min2}
\end{equation}
As in the previous case we can obtain the mass matrices for the
scalar fields starting from the potential in Eq. (\ref{V2}). 
For example, using Eq. (\ref{min2}) one obtains for the neutral pseudoscalars 
$(\chi_1 ~\chi_2 ~\chi_3)$:
\begin{equation}
 M^2_{\chi_i \chi_j} =  \left(\frac{v^2}{4}\right) 
\pmatrix{ -2\lambda_4  & 2 \lambda_4  & 0 \cr 
 2 \lambda_4  & -2 \lambda_4  & 0  \cr 0 & 0 & 
- 3\lambda_2/4 + \lambda_3/4 -3 \lambda_4 } \;.
\label{chii2}
\end{equation}
Similarly, for the real scalars $(\phi_1 ~\phi_2 ~\phi_3)$ one has:
\begin{equation}
 M^2_{\phi_i \phi_j} = \left(\frac{v^2}{4}\right)
 \pmatrix{ \lambda_1 + \lambda_2  & 
\lambda_1 - \lambda_2/2 + \lambda_3/2 + 2 \lambda_4  & 0 \cr 
\lambda_1 - \lambda_2/2 + \lambda_3/2 + 2 \lambda_4  & 
\lambda_1 + \lambda_2  & 0  \cr 0 & 0 & 
- 3\lambda_2/4 + \lambda_3/4 + \lambda_4 } \;.
\label{phii2}
\end{equation}

The charged scalar mass matrix is found to be:
\begin{equation}
M^2_{\phi^\mp_i \phi^\pm_j} = \left(\frac{v^2}{4}\right)
\pmatrix{-\lambda_3/4 - \lambda_4  & \lambda_3/4 + \lambda_4 & 0 \cr 
\lambda_3/4 + \lambda_4 & -\lambda_3/4 - \lambda_4   & 0 \cr 0 &
0 & -3\lambda_2/4 - \lambda_3/4 - \lambda_4} \;.
\label{chgd2}
\end{equation}
 
To go to the Higgs basis using Eq. (\ref{Higgsbasis}) one
must use a unitary transformation by:
\begin{equation}
U_2 = \pmatrix{1/\sqrt{2} & 1/\sqrt{2} & 0 \cr
-1/\sqrt{2} & 1/\sqrt{2} & 0 \cr
0 & 0& 1} \;.
\end{equation}
The mass matrices  in Eqs. (\ref{chii2}) - (\ref{chgd2}) are all
diagonalised by the same unitary transformation $U_2 M^2 U_2^\dagger
= D^2$ where $D$ is diagonal. We find: 
\begin{equation}
 D^2_{\chi_i \chi_j} = \left(\frac{v^2}{4}\right) 
{\rm diag} (0, ~-4 \lambda_4, ~- 3\lambda_2/4 + \lambda_3/4 -3 \lambda_4 ) \;, 
\label{chii2a}
\end{equation}
\begin{equation}
 D^2_{\phi_i \phi_j} =  \left(\frac{v^2}{4}\right)
 {\rm diag} (2 \lambda_1 + \lambda_2/2 + \lambda_3/2 + 2 \lambda_4 , 
~3\lambda_2/2 - \lambda_3/2 - 2 \lambda_4, 
~- 3\lambda_2/4 + \lambda_3/4 +  \lambda_4 ) \;,
\label{phii2a}
\end{equation}
and
\begin{equation}
 D^2_{\phi^\mp_i \phi^\pm_j} = \left(\frac{v^2}{4}\right) 
  {\rm diag} (0, ~-\lambda_3/2 - 2\lambda_4, ~-3\lambda_2/4 -
\lambda_3/4 - \lambda_4)  \;.
\label{chgd2a}
\end{equation}

It is important to note that we are defining the Higgs basis through:
\begin{equation}
\Psi = U_2 \Phi \;,
\label{basistr}
\end{equation}
where $\Phi$ is given in Eq. (\ref{multipletp})  and $\Psi$
is defined in Eq. (\ref{Higgsbasis}).

In this Higgs basis the scalars are mass eigenstates as expected,
where $\psi_1^+$ and $\xi_1$ are massless Goldstones and $\eta_1$
is massive. Further, the vev is:
\begin{equation}
\langle \Psi \rangle_2 = \frac{v}{\sqrt{2}} \pmatrix{0 & 1 \cr 0
& 0 \cr 0 & 0}  \;,
\label{vev2a}
\end{equation}
which makes the alignment obvious.

Notice that the second and third eigenvalues of Eq.
(\ref{phii2a}) are proportional but with opposite sign. So, both
cannot be made positive by any choice of the $\lambda_i$. This is
an inadequacy which can be removed by choosing $\lambda_4$ to be
complex, where alignment still continues to be valid. We
demonstrate this in an Appendix.

\subsection{Case 3: $\langle\phi_i^0\rangle = \frac{v}{\sqrt{6}} (1,1,1)$}

Next we consider $\langle\phi_1^0\rangle = \langle\phi_2^0\rangle =
\langle\phi_3^0\rangle = {v \over \sqrt{6}}$, i.e., 
\begin{equation}
\langle \Phi \rangle_3 = {v \over \sqrt{6}} \pmatrix{0 & 1 \cr 0
& 1 \cr 0 & 1}  \;.
\label{vev3}
\end{equation}

In this case the minimisation of the potential implies:
\begin{equation}
m^2 + \frac{v^2}{12} \left[3\lambda_1  + \lambda_3 +
4\lambda_4\right] = 0 \;.
\label{min3}
\end{equation}
Using the above one can calculate the mass matrices of the scalar
fields. For example, for the neutral pseudoscalars 
$(\chi_1 ~\chi_2 ~\chi_3)$ one gets:
\begin{equation}
 M^2_{\chi_i \chi_j} = 2 \lambda_4 \left(\frac{v^2}{6}\right) 
\pmatrix{-2 & 1 & 1 \cr 1 & -2 & 1 \cr 1 & 1 & -2 } \;.
\label{chii3}
\end{equation}
The real scalar $(\phi_1 ~\phi_2 ~\phi_3)$ mass matrix in this case is:
\begin{equation}
 M^2_{\phi_i \phi_j} =  \left(\frac{v^2}{6}\right) 
\pmatrix{y & z & z \cr z & y & z \cr z & z & y } \;,
\label{phii3}
\end{equation}
where $y = (\lambda_1 + \lambda_2)$ and $z = (\lambda_1 -
\lambda_2/2 + \lambda_3/2 + 2\lambda_4)$.

Finally, for the charged sector
\begin{equation}
M^2_{\phi^\mp_i \phi^\pm_j} = \left(\frac{v^2}{6}\right)
 \left(\lambda_4 +\frac{\lambda_3}{4}\right) 
\pmatrix{-2 & 1 & 1 \cr 1 & -2 & 1 \cr 1 & 1 & -2 } \;.
\label{chgd3}
\end{equation}

The above mass matrices are all diagonalised by a unitary
transformation by the matrix $U_3$ defined in Eq. (\ref{u3}).
Thus, $U_3$ rotates the defining basis to the Higgs
basis\footnote{Notice that in all three scalar sectors there is
double degeneracy and consequently the Higgs basis is non-unique.
For example, in place of $U_3$ of Eq. (\ref{u3}) one could just as
well use the popular tribimaximal mixing matrix: $U_{TBM} =
\pmatrix{\frac{1}{\sqrt{3}} &
\frac{1}{\sqrt{3}} & \frac{1}{\sqrt{3}} \cr
-\frac{\sqrt{2}}{\sqrt{3}} & \frac{1}{\sqrt{6}} &
\frac{1}{\sqrt{6}} \cr 0 & -\frac{1}{\sqrt{2}} & \frac{1}{\sqrt{2}}}$.}.

The diagonal forms of the mass matrices are:
\begin{equation}
 D^2_{\chi_i \chi_j} = \lambda_4 ~v^2 
{\rm diag} (0, ~-1, ~-1) \;, 
\label{chii3a}
\end{equation}
\begin{equation}
 D^2_{\phi^\mp_i \phi^\pm_j} = 
 \left(\lambda_4 +\frac{\lambda_3}{4}\right) \left(\frac{v^2}{2}\right)
  {\rm diag} (0, ~-1, ~-1) \;,
\label{chgd3a}
\end{equation}
and
\begin{equation}
 D^2_{\phi_i \phi_j} =  \left(\frac{v^2}{6}\right)
 {\rm diag} (y + 2 z, ~y - z , ~y - z) \;,
\label{phii3a}
\end{equation}
where $y + 2z = 3 \lambda_1 +  \lambda_3 + 4 \lambda_4$ and $y
- z = 3 \lambda_2/2 - \lambda_3/2 - 2 \lambda_4$. Both these
combinations can be positive while remaining consistent with
positivity and unitarity. Similarly, $\lambda_4$ and $\lambda_4 +
\lambda_3/4$, which appear in Eqs. (\ref{chii3a}) and
(\ref{chgd3a}) respectively,  can both be negative.   

The fields in the Higgs basis are:
\begin{equation}
\Psi = U_3 \Phi \;,
\label{basistr3}
\end{equation}
where $\Phi$ is given in Eq. (\ref{multipletp}).  As before, we
use $\psi_i^0 = \eta_i + i \xi_i$.

Further, in this basis in which the scalars are mass eigenstates,
with $\psi_1^+$ and $\xi_1$ massless, the
vev becomes
\begin{equation}
\langle \Psi \rangle_3 = \frac{v}{\sqrt{2}} \pmatrix{0 & 1 \cr 0
& 0 \cr 0 & 0}  \;.
\label{vev3a}
\end{equation}
Thus, alignment is again manifest.

\subsection{Case 4: $\langle\phi_i^0\rangle = \frac{v}{\sqrt{6}}  (1, ~\omega, ~\omega^2)$} 

The last alternative that we consider involves complex vacuum
expectation values, namely,  $\langle\phi_1^0\rangle = (v/\sqrt{6})$, $\langle\phi_2^0\rangle =
(v/\sqrt{6}) \omega$, and $\langle\phi_3^0\rangle = (v/\sqrt{6}) \omega^2$, i.e., 
\begin{equation}
\langle \Phi \rangle_4 = {v \over \sqrt{6}} \pmatrix{0 & 1 \cr 0
& \omega \cr 0 & \omega^2}  \;.
\label{vev4}
\end{equation}
The vev is in this direction if
\begin{equation}
m^2 + \frac{v^2}{12} \left[3\lambda_1  + \lambda_3 -
2\lambda_4\right] = 0 \;.
\label{min4}
\end{equation}

Since the vev are complex there will be mixing terms involving
neutral scalars and pseudoscalars. The $(6 \times 6)$ mass matrix
in the $(\chi_1, \chi_2, \chi_3, \phi_1, \phi_2, \phi_3)$ basis is:
\begin{equation}
 M^2_{\Phi^0_i \Phi^0_j} = \frac{v^2}{6}  
\pmatrix{2\lambda_4 & -\lambda_4  &  -\lambda_4
& 0 & \sqrt{3} \lambda_4 &   -\sqrt{3} \lambda_4 \cr -\lambda_4
& f_2  & f_1 & g_1 & g_2 &  g_3 \cr -\lambda_4 &  f_1 &  
f_2 & -g_1 & -g_3 & -g_2 \cr
0 & g_1  & -g_1 & (\lambda_1 + \lambda_2) 
& h_1 & h_1 \cr  \sqrt{3} \lambda_4 & 
g_2 & -g_3 & h_1 & h_3 & h_2
\cr  -\sqrt{3} \lambda_4 & g_3 & -g_2 & h_1 & h_2 & h_3} \;,
\label{chii4}
\end{equation}
where
\begin{eqnarray}
f_1 &=& -\frac{3}{4} \left[\lambda_1 - \lambda_2/2 +
\lambda_3/2 \right] - \lambda_4  \;,\;
f_2 = \frac{3}{4}(\lambda_1 + \lambda_2) + \frac{1}{2} \lambda_4 \;,\nonumber \\ 
g_1 &=& \frac{\sqrt{3}}{4}\{2\lambda_1  - \lambda_2  +
\lambda_3 - 4\lambda_4 \} \;,
\; g_2 = -\frac{\sqrt{3}}{4}\{\lambda_1 + \lambda_2 - 2 \lambda_4\}  \;,\;
g_3 = -\frac{\sqrt{3}}{4}\{\lambda_1  - \lambda_2/2  +
\lambda_3/2 - 4\lambda_4 \} \;,  \nonumber \\
h_1 &=& -\frac{1}{4}\{2\lambda_1 - \lambda_2  + \lambda_3 
+ 4 \lambda_4\} \; , \;
h_2 = \frac{1}{8}\{2\lambda_1 - \lambda_2  + \lambda_3 
-8 \lambda_4\} \; , \;
h_3 = \frac{1}{4}\{\lambda_1 + \lambda_2 + 6 \lambda_4\} \;.
\label{symbol}
\end{eqnarray}
The matrix in Eq. (\ref{chii4}) has an eigenstate with zero
eigenvalue. This state can be readily identified by changing the
basis through  a $(6 \times 6)$ unitary transformation by
\begin{equation}
U_{6r} = \frac{1}{\sqrt{3}}\pmatrix{1 & 1 & 1 & 0 & 0 & 0 \cr 1 & -1/2 & -1/2 & 0 &
-\sqrt{3}/2 & \sqrt{3}/2 \cr 1 & -1/2 & -1/2 & 0 & \sqrt{3}/2 &
-\sqrt{3}/2 \cr 0 & 0 & 0& 1 & 1& 1 \cr 0 & \sqrt{3}/2 &
-\sqrt{3}/2 & 1 & -1/2 & -1/2 \cr 0 & -\sqrt{3}/2 & \sqrt{3}/2 &
1 & -1/2 & -1/2} \;.
\label{u6}
\end{equation}
The new basis thus obtained is:
\begin{equation}
\pmatrix{\xi_1 \cr \xi_2 \cr \xi_3 \cr \eta_1 \cr \eta_2 \cr
\eta_3} = \frac{1}{\sqrt{3}} \pmatrix{\chi_1 + \chi_2 + \chi_3 \cr 
\chi_1 - (\chi_2 + \chi_3)/2 - \sqrt{3}(\phi_2 - \phi_3)/2 \cr
\chi_1 - (\chi_2 + \chi_3)/2 + \sqrt{3}(\phi_2 - \phi_3)/2 \cr
\phi_1 + \phi_2 + \phi_3 \cr
 \sqrt{3}(\chi_2 - \chi_3)/2 + \phi_1 - (\phi_2 + \phi_3)/2  \cr
- \sqrt{3}(\chi_2 - \chi_3)/2 + \phi_1 - (\phi_2 + \phi_3)/2 } \;.  
\end{equation}
It turns out that $\xi_2$ and $\eta_2$ are mass eigenstates with
$\xi_2$ being the mass-zero mode. The rest of the
mass matrix separates into two block diagonal forms, a $(2 \times
2)$ block  for $(\xi_1, \xi_3)$ and another for 
$(\eta_1, \eta_3)$ which are:
\begin{equation}
M^2_{\xi_1,\xi_3} = \frac{v^2}{6}\pmatrix{
\lambda_A &
-\lambda_A \cr -\lambda_A & \lambda_A +  18 \lambda_4} \;,\;
M^2_{\eta_1, \eta_3} = \frac{v^2}{6}\pmatrix{
\lambda_A  & \lambda_A \cr \lambda_A &  \lambda_A 
+ 18 \lambda_4} \;,
\label{mass22}
\end{equation}
where $\lambda_A = \frac{9}{4} \lambda_2 - \frac{3}{4} \lambda_3
- 3 \lambda_4$. $m^2_{\xi_2} = 0$ and $m^2_{\eta_2} =
v^2 ~(3 \lambda_1/2 + \lambda_3/2 -  \lambda_4)$.  The
eigenvalues and eigenvectors of the two matrices in Eq.
(\ref{mass22}) are: 
\begin{eqnarray}
m_{1}^2 &=& \frac{v^2}{6} \left[\lambda_A + 9 \lambda_4 +
\sqrt{\lambda_A^2 + 81 \lambda_4^2} \right] \;,\; \xi_1 = 
\chi_1 \cos\alpha - \chi_3 \sin \alpha \;,\; \eta_1 = \phi_1 \cos
\alpha + \phi_3  \sin \alpha \nonumber \\
m_{3}^2 &=& \frac{v^2}{6} \left[\lambda_A + 9 \lambda_4 -
\sqrt{\lambda_A^2 + 81 \lambda_4^2} \right] \;,\; \xi_3 = 
 \chi_1 \sin \alpha + \chi_3 \cos \alpha \;,\; \eta_3 = -\phi_1 \sin
\alpha + \phi_3  \cos \alpha \;,
\end{eqnarray}
where
\begin{equation}
\tan 2\alpha = \frac{\lambda_A}{9 \lambda_4} \;.
\end{equation}

At this stage we draw attention to the fact that the $(6 \times
6)$ unitary matrix, $U_{6r}$ in Eq. (\ref{u6}), acting on real fields
$(\chi_i, \phi_i)$  is nothing but a unitary transformation by $U_3$ of
Eq. (\ref{u3}) on the complex fields, $\phi_i^0 = \phi_i + i \chi_i$.

For the charged sector (after eliminating $\lambda_{1,2}$ using
Eq. (\ref{min4})):
\begin{equation}
 M^2_{\phi^\mp_i \phi^\pm_j} = \frac{v^2}{6}  
\pmatrix{a & b & b^* \cr b^* & a & b \cr b & b^* & a } \;,
\label{chgd4}
\end{equation}
where $a = (2\lambda_4 - \lambda_3)/2$ and $b = (\omega^2
\lambda_3 + 4\omega \lambda_4)/4$.  This matrix is also
diagonalised by going to the $\Psi$ basis using Eq.
(\ref{Higgsbasis}) with $U_3$ from Eq. (\ref{u3}) and one has the
eigenvalues $(a + 2 Re(b)) = - \frac{v^2}{6} (3 \lambda_3/4)$,
$(a - Re(b) -\sqrt{3} Im(b)) = 0 $, and $(a - Re(b) + \sqrt{3}
Im(b)) = -\frac{v^2}{6}(3 \lambda_3/4 - 3 \lambda_4)$. The
corresponding eigenstates are precisely:
\begin{equation}
\psi^\pm_1 = (\phi^\pm_1 + \phi^\pm_2 + \phi^\pm_3)/\sqrt{3} \;,\; 
\psi^\pm_2 = (\phi^\pm_1 + \omega \phi^\pm_2 + \omega^2
\phi^\pm_3)/\sqrt{3} \;,\;
\psi^\pm_3 = (\phi^\pm_1 + \omega^2 \phi^\pm_2 + \omega
\phi^\pm_3)/\sqrt{3} \;.
\end{equation} 
Thus $(\Psi_1, ~\Psi_2, ~\Psi_3)$ constitute the Higgs basis for this
case with $\Psi_2$ mimicing the Standard Model scalar doublet in this case.

Note, that in this Higgs basis the vev becomes:
\begin{equation}
\langle \Psi \rangle_4 = \frac{v}{\sqrt{2}} \pmatrix
{0 & 0 \cr 0
& 1 \cr 0 & 0}  \;.
\label{vev4v2}
\end{equation}

The fact that in the $\Psi$-basis the vev takes the form in Eq.
(\ref{vev4v2}) and that the components of $\Psi_2$, namely
$(\psi^+_2 , \psi^0_2)$, are both mass
eigenstates with massless charged and neutral modes and a massive
neutral scalar exemplifies
alignment in this case.

\section{Positivity, Unitarity}

The scalar potential in Eq. (\ref{V2}) must be bounded from
below. This gives rise to `positivity' bounds on the couplings
appearing in it. Further, tree-level unitarity of scalar-scalar
scattering also gives rise to bounds on combinations of the same
couplings. In the following we show that these constraints still do
permit the $\lambda_i$ to be chosen such that the mass-squareds of
all scalars are either positive or vanishing.

\subsection{Positivity limits}
It is well-known that if the scalar potential of any model
depends only on the squares of the fields then one has to
consider the `copositivity' constraints on the couplings. For a
general model with three scalar doublets such constraints are
available in the literature \cite{kannike, joydeep}. We adopt
these to the model with $A4$ symmetry. 

Because of the $A4$ symmetry, the scalar quartic couplings are
few and related. If all vacuum expectation values are real one
has to look for the copositivity of the matrix
\begin{equation}
M_{cop} = \pmatrix{\lambda_P  & \lambda_Q & \lambda_Q \cr
\lambda_Q & \lambda_P & \lambda_Q \cr
\lambda_Q & \lambda_Q & \lambda_P } \;,
\label{copos}
\end{equation}
where the combinations
\begin{equation}
\lambda_P = (\lambda_1 + \lambda_2)/2 \;,\; \lambda_Q = 
(2\lambda_1 - \lambda_2 + \lambda_3 + 4 \lambda_4)/4 \;.
\end{equation}
The conditions to be satisfied are:
\begin{equation}
\lambda_P \geq 0 \;,\;
\lambda_P + \lambda_Q \geq 0 \;,\; {\rm and} \;
\sqrt{\lambda_P^3} + (3 \lambda_Q) ~\sqrt{\lambda_P}  +
\sqrt{2(\lambda_P + \lambda_Q)^3} ~\geq 0 \;. 
\label{pos1}
\end{equation}  
To satisfy these conditions it is enough to demand $\lambda_P
\geq 0$ and $\lambda_Q \geq -\frac{1}{2} \lambda_P$ which
translate to:
\begin{equation}
\lambda_1 + \lambda_2 \geq 0 \;,\; 3\lambda_1 + \lambda_3 + 4
\lambda_4 \geq 0 \;.
\label{pos0}
\end{equation}

For complex vev, in general, one has to look at the {\em
positivity} of the matrix. However, in the simpler situation in
the Case 4 discussed earlier, where the phases of $\langle \Phi_1
\rangle$, $\langle\Phi_2 \rangle$ and $\langle \Phi_3 \rangle$
are $(0, 2\pi/3, 4\pi/3)$, the copositivity criteria continue to
apply. The matrix $M_{cop}$ is the same as in Eq. (\ref{copos})
except for
\begin{equation}
\lambda_Q \rightarrow \lambda_R = (2\lambda_1 - \lambda_2 +
\lambda_3 -2 \lambda_4)/4 \;.
\label{pos2}
\end{equation}
In this case, one must satisfy 
\begin{equation}
\lambda_1 + \lambda_2 \geq 0 \;,\; 3\lambda_1 + \lambda_3 - 2
\lambda_4 \geq 0 \;.
\label{pos3}
\end{equation}

\subsection{$s$-wave unitarity}

The potential in Eq. (\ref{V2}) involves $SU(2)_L$ doublet scalar
fields ($I = 1/2$) with $Y = 1$ and their hermitian conjugates.
The quartic terms in the potential can give rise to tree-level
scalar-scalar scattering processes. At high energies one can
classify the  scattering states by their $SU(2)_L$ and $Y$
quantum numbers.  The two-particle states can be in $SU(2)_L$
singlet ($I = 0$) or $SU(2)_L$ triplet ($I = 1$) channels and for
both cases with $Y$ =  2 (e.g., $\phi_i^+\phi_j^+$ initial/final
states, $i,j = 1,2,3$) or 0 (e.g., $\phi_i^+\phi_j^{*0}$
initial/final states, $i,j = 1,2,3$). The scattering processes
in every channel must respect limits arising from
probabilitiy conservation, i.e., unitarity. This implies bounds
on the amplitudes for each partial wave. Here we restrict
ourselves to the bounds from $s$-wave scattering  for the
different channels \cite{2hdmu, unitary2}.

A discussion of the unitarity bounds for the two-scalar-doublet
model along these lines can be found in \cite{2hdmu}. It can be
readily generalised to the three-scalar-doublet case under
consideration here.

The results we obtained are displayed in Table \ref{t1}.  Besides
$I$ and $Y$ an initial or final state will carry two indices
$(i,j)$ when the two scalars are from  $\Phi_i$ and $\Phi_j$. We
treat two cases separately. `Diagonal' corresponds to states with
$i = j$ while `Off-diagonal' is for $i \neq j$.  Since in all
terms in the potential in Eq. (\ref{V2}) any field, $\Phi_i$,
appears an even number of times, a `Diagonal' initial state
cannot scatter into an `Off-diagonal' state and {\em vice versa}.
So, the two sectors are completely decoupled for the
$A4$-symmetric potential.  Note that for the `Diagonal' case
there is no  $SU(2)_L$ singlet state with $Y = 2$ due to Bose
statistics.

\begin{table}[tb]
\begin{center}
\begin{tabular}{|c|c|c|c| c|}
\hline
\multicolumn{2}{|c|}{Quantum numbers}& Type  & Matrix &
Eigenvalues \\ \cline{1-2}
$SU(2)_L$ & $Y$ &  & size &  \\ \hline
1 & 2 & Diagonal & $3 \times 3$ & $|(\lambda_1 - 2 \lambda_4)| \;,\;
|(\lambda_1 + 4 \lambda_4)|$   \\ \hline
1 & 2 & Off-diagonal  &  $3 \times 3$ & $|(\lambda_3 - 3 \lambda_2)/2| $   \\ \hline
0 & 2 & Off-diagonal  &  $3 \times 3$ & $|(\lambda_3 + 3 \lambda_2)/2| $   \\ \hline
\hline
1 & 0 & Diagonal &  $3 \times 3$ & $|(\lambda_1 -  \lambda_3/2)| \;,\;
|(\lambda_1 + \lambda_3)|$   \\ \hline
1 & 0 & Off-diagonal  &  $6 \times 6$ & $|(3\lambda_2 + 4
\lambda_4)/2| \;,\;
|(3\lambda_2 - 4 \lambda_4)/2|$   \\ \hline
0 & 0 & Diagonal &  $3 \times 3$ &  $|(6\lambda_1 +
6\lambda_2- \lambda_3)/2|
\;,\; |(3\lambda_1 - 6 \lambda_2 + \lambda_3)|$   \\ \hline
0 & 0 & Off-diagonal  &  $6 \times 6$ & $|(-3\lambda_2 + 2
\lambda_3 - 12 \lambda_4)/2|
\;,\; |(-3\lambda_2 + 2 \lambda_3 + 12 \lambda_4)/2|$  \\ \hline
\end{tabular}
\end{center}
\caption{\sf The dimensionalities and eigenvalues of the tree-level
scattering matrices for the different $SU(2)_L$ and $Y$ sectors.
`Diagonal' (`Off-diagonal') corresponds to $i = j$ ($i \neq j$)
with $i,j = 1,2,3$. From unitarity the magnitude of each
eigenvalue must be bounded by 1/8$\pi$.}
\label{t1}
\end{table}

We denote the $s$-wave scattering amplitude by $S(I,Y)$.  For
each choice of $I, ~I_3$, and $Y$ quantum numbers there are a
fixed set of states determined by the  available options for
$i$ and $j$. A scattering matrix can be obtained for an initial
state from this set going over to a final state also from this
set.  $s$-wave unitarity requires every eigenvalue of the matrix
to be bounded by $1/8\pi$. For example, for the $S(0, 2)$
`Off-diagonal' case the states are $\Gamma_{12}(0, 2) \equiv
(\phi_1^+\phi_2^0 - \phi_1^0\phi_2^+)/\sqrt{2}$, $\Gamma_{23}(0,
2) \equiv (\phi_2^+ \phi_3^0 - \phi_2^0\phi_3^+)/\sqrt{2}$,
$\Gamma_{31}(0, 2) \equiv (\phi_3^+ \phi_1^0 -
\phi_3^0\phi_1^+)/\sqrt{2}$.  Scattering between these
states\footnote{Notice that $\Gamma_{ij}(0,2)
=-\Gamma_{ji}(0,2)$.} gives rise to a $(3 \times 3)$ matrix.
The matrix is diagonal because no term in the potential in
Eq. (\ref{V2}) can cause an off-diagonal transition in this
sector, e.g., $\Gamma_{12}(0, 2) \not\leftrightarrow
\Gamma_{23}(0, 2)$,  which involves an odd number of $\Phi_1$ and
$\Phi_3$ fields. Also, the matrix is proportional to the identity
due to the $A4$ symmetry.  On the other hand, for the $S(0,0)$
`Off-diagonal' case for any $i$ and $j  \; (i \neq j)$ the
initial and final states can be any one of $(\phi_i^+\phi_j^- +
\phi_i^0\phi_j^{*0})/\sqrt{2}$ or $(\phi_i^-\phi_j^+ +
\phi_i^{*0}\phi_j^0)/\sqrt{2}$. This results in a $(6 \times 6)$
matrix which turns out to be of block diagonal form with three
identical $(2 \times 2)$ blocks.

In Table \ref{t1} we have listed the different channels, the
corresponding scattering matrix dimensions, and their
eigenvalues.   We present below two typical examples of the
matrices, corresponding to the first and fifth rows of Table
\ref{t1}.
\begin{equation}
8 \pi S(1,2)_{diag} = \pmatrix{\lambda_1 & 2 \lambda_4 & 2
\lambda_4 \cr 2 \lambda_4 & \lambda_1 & 2 \lambda_4 \cr 2
\lambda_4 & 2 \lambda_4 & \lambda_1} ,
8 \pi S(1,0)_{off-diag} = \pmatrix{X & 0 & 0 \cr 0 & X & 0 \cr 0 &
0 & X} ,  X = \pmatrix{ -3\lambda_2/2 & 2 \lambda_4 \cr
2 \lambda_4 & -3\lambda_2/2}
 \;.
\end{equation}
For all cases at least one (or more) of the
eigenvalues is degenerate due to the $A4$ symmetry requirement on
the potential.

Note that the above discussion of tree-level unitarity has been in
terms of the defining scalar fields $\Phi_{1,2,3}$. An alternate
approach which also appears in the literature is to consider the
unitarity bounds following from the scattering of physical mass
eigenstate scalars.  The results are equivalent.

\subsection{Scalar mass-squareds}

In section \ref{vevcases} we have obtained the mass eigenvalues
of the charged and neutral scalars for the four alternate
choices of the vevs. We find from the limits on the quartic
couplings from positivity and from unitarity (Table
\ref{t1})  that there is ample room to choose the $\lambda_i$
such that all scalar mass-squareds are positive, i.e., all masses
are real.
 
In particular, positivity of the mass-squared for all the physical
fields requires: 

\[
{\rm Case ~1:}  ~\lambda_1 + \lambda_2 > 0 \;,\; \lambda_2 < 0
\;,\; 3 \lambda_2 - \lambda_3 + 4 \lambda_4 < 0 \;,\; 
3 \lambda_2 - \lambda_3 - 4 \lambda_4 < 0 \;.
\]
Of the above conditions, the first is a requirement that must be
met, see Eq. (\ref{pos0}), for the potential to be positive.
Besides, one must make the choices $\lambda_2 <0$ and $(3\lambda_2
- \lambda_3 + 4 |\lambda_4|) <0$. 
\[
{\rm Case ~2:  ~As ~noted ~in ~section ~\ref{case2},  ~for
~real} ~\lambda_4 {\rm ~this ~case ~is ~inadmissible.}
\]
In the Appendix we show that this issue is removed if $\lambda_4$
is taken complex.  
\[
{\rm Case ~3:}  ~\lambda_4 < 0 \;,\; \lambda_3 + 4 \lambda_4 < 0 \;,\; 
3 \lambda_1 + \lambda_3 + 4 \lambda_4 > 0 \;,\; 3\lambda_2 -
\lambda_3 - 4 \lambda_4 > 0 \;. 
\]
The third inequality in the above is again a consequence of the
positivity of the potential, Eq. (\ref{pos0}). Along with this,
the imposition of the first and second conditions imply the
fourth.
\[
{\rm Case ~4:}  ~3 \lambda_1 + \lambda_3 -2 \lambda_4 > 0 \;,\;
\lambda_3 < 0 \;,\; \lambda_3 - 4 \lambda_4 < 0 \;,\; \lambda_4 ~(3\lambda_2 -
\lambda_3 - 4 \lambda_4) > 0 \;. 
\]
The first condition is anyway satisfied for the positivity of the
potential, see Eq. (\ref{pos3}). Further, this soultion is viable
in the parameter region defined by the remaining conditions.

The constraints from unitarity in Table \ref{t1} set bounds on
the magnitude of some combinations of these couplings. Barring
unnatural cancellations, this implies that none of the $\lambda_i$ can be
arbitrarily large in magnitude.

\section {Conclusions}

In this paper we have considered a three-Higgs doublet model with
$A4$ symmetry.  We have examined the vevs which have been
identified as the {\em only} possible global minima of this potential. These
choices have been used in realistic physics models. Here we have
shown that in all these cases alignment is automatic due to the
$A4$ symmetry. We demonstrate that the bounds on the quartic couplings
from positivity and unitarity can be satisfied while keeping all
scalar mass-squareds positive.

The attractive feature of $A4$ symmetry is that the terms
that are allowed in the potential ensure distinctive textures for
the scalar mass matrices.  Alignment for the global minima vevs 
is a result of this. Another consequence is that
the mixing matrices in the scalar sector are of a few standard
forms.

Needless to say, since alignment is valid, with the minimal
scalar content considered in this work, in the Higgs basis
fermion masses will arise from the coupling to the analogue of
the SM Higgs boson in this model. Usually, however, realistic
models involve the inclusion of more scalar fields.  Such a model
based on $A4$ symmetry incorporating fermions and reproducing
their observed mass and mixing patterns through Yukawa couplings
is beyond the scope of this work. 

In conclusion, this model exhibits several features which make it
of interest for further exploration.

{\bf Acknowledgements:} SP acknowledges support from CSIR, India
through a NET Senior Research Fellowship.
AR is partially funded by  SERB Grant No. SR/S2/JCB-14/2009 and
SERB grant No. EMR/2015/001989.

\renewcommand{\thesection}{\Alph{section}} 
\setcounter{section}{0} 
\renewcommand{\theequation}{\thesection.\arabic{equation}}  
\setcounter{equation}{0}  

\section{Appendix: Case 2 generalisation}\label{App}

It was shown in sec. \ref{case2} that though  alignment is valid
for the vev $\langle \Phi_0\rangle = (v/2)(1, 1, 0)$  the
physical scalar mass-squareds cannot all be made simultaneously
positive when the coupling $\lambda_4$ is real. We show in this
Appendix that this issue is addressed when
$\lambda_4$ is complex.
 
We take $\lambda_4 = |\lambda_4| e^{i \delta}$ and the vev of the
neutral scalars as $(v/2)(1, ~e^{i \alpha}, ~0)$. In this case,
the minimisation condition becomes:
\begin{equation}
m^2 + \frac{v^2}{4}\left[\lambda_1 + \frac{1}{4} \lambda_2 +
\frac{1}{4} \lambda_3 - |\lambda_4|\right] = 0 \; {\rm and} \; \delta +
2 \alpha = \pi
\end{equation}
Note that the phase of $\lambda_4$ is related to that of the vev.

If $\lambda_4$ is complex then the charged scalar sector
$(\phi_1^\pm, ~\phi_2^\pm, \phi_3^\pm)$ 
mass-squared matrix is:
\begin{equation} M^2_{\phi^\mp_i \phi^\pm_j} =
\left(\frac{v^2}{4}\right)
\pmatrix{-\lambda_3/4 + |\lambda_4|  & e^{-i\alpha}(\lambda_3/4 -
|\lambda_4|) & 0 \cr e^{i\alpha}(\lambda_3/4 - |\lambda_4|) &
-\lambda_3/4 + |\lambda_4|   & 0 \cr 0 & 0 & -3\lambda_2/4
-\lambda_3/4 + |\lambda_4| } \;.
\label{chgd2b}
\end{equation}
This matrix is diagonalised by a unitary transformation of the fields:
\begin{equation}
\pmatrix{\Psi_1 \cr \Psi_2 \cr \Psi_3} = U_{2c}
\pmatrix{\Phi_1 \cr \Phi_2 \cr \Phi_3} \; {\rm ~with} \;
U_{2c} = \frac{1}{\sqrt{2}} \pmatrix{1 & e^{-i\alpha} & 0
\cr e^{i\alpha}  & -1 & 0 \cr 0 & 0 &  1}  \;.
\label{basiscomp}
\end{equation}
As before, we write $\Psi_i^0 = \eta_i + i \xi_i$. Notice that
in this basis the vev is of the form $\langle\Psi^0\rangle =
(v/\sqrt{2}) (1, 0, 0)$.

The charged states $(\psi_1^\pm, \psi_2^\pm, \psi_3^\pm)$ have
masses $0, ~(v/2)\sqrt{-\lambda_3/2 + 2 |\lambda_4|}, ~(v/2)\sqrt{-3\lambda_2/4 -
\lambda_3/4 +  |\lambda_4|}$ respectively. That $\psi_1^\pm$ is
massless is indicative that the $\Psi_i$ constitute the Higgs
basis. To establish alignment we need to check that $\xi_1$ is
massless and $\eta_1$ has a positive mass-squared.

Because the vev and $\lambda_4$ are now complex there is
scalar-pseudoscalar mixing in the neutral sector. The neutral
sector mass-squared matrix splits up into a $(2\times 2)$ block
for $(\chi_3, \phi_3)$ which remains decoupled from the remaining
$(4 \times 4)$ block. For the other neutral fields, i.e.,  $(\chi_1, \chi_2,
\phi_1, \phi_2)$ states:
\begin{equation}
M^2_{\chi_1, \chi_2, \phi_1, \phi_2} = \frac{v^2}{4}
2 |\lambda_4| \left[ I + \pmatrix{0 & -\cos\alpha & 0 & \sin\alpha
\cr  -\cos\alpha & K \sin^2\alpha  &  J \sin\alpha &
K \sin\alpha \cos\alpha  \cr 0 & J \sin\alpha  & K
&  J \cos\alpha  \cr \sin\alpha & K \sin\alpha \cos\alpha 
&  J \cos\alpha & K \cos^2\alpha } \right] \;.
\label{neutral4a}
\end{equation}
Here $K = (\lambda_1 + \lambda_2 - 2|\lambda_4|)/2|\lambda_4|$
and $J = (\lambda_1 - \lambda_2/2 + \lambda_3/2 -
2|\lambda_4|)/2|\lambda_4|$.

This matrix is diagonalised by a $(4 \times 4)$ unitary matrix
which is the upper $(2
\times 2)$ block of $U_{2c}$ in Eq. (\ref{basiscomp})
expressed in this $(\chi_{1,2}, ~\phi_{1,2})$ basis, i.e.,:
\begin{equation}
U_{4r} = \frac{1}{\sqrt{2}}\pmatrix{1 & \cos\alpha & 0 & -\sin\alpha
\cr  \cos\alpha & -1 &  \sin\alpha & 0 \cr 0 & \sin\alpha & 1 &
\cos\alpha \cr -\sin\alpha & 0 &  \cos\alpha &  -1 } \;.
\label{u4}
\end{equation}
We find 
\begin{equation}
U^\dagger_{4r} ~[M^2_{\chi_1, \chi_2, \phi_1, \phi_2}] ~U_{4r} =
\frac{v^2}{4} 2 |\lambda_4| \pmatrix{0 & 0 & 0 & 0 \cr  0 & 2 +
(-1 + K - J)\sin^2\alpha  &  0 & (-1 + K - J) \sin\alpha
\cos\alpha  \cr 0 & 0 & (1+K + J) &  0 \cr 0 & (-1 + K - J)
\sin\alpha \cos\alpha  &  0 & 2 + (-1 + K - J) \cos^2\alpha  }
\;.
\label{neutral4}
\end{equation}
Thus, as required for alignment, the masses of $\xi_1$ and
$\eta_1$ are:
\begin{equation}
m_{\xi_1} = 0 \;,\; m_{\eta_1} = \frac{v}{2}\sqrt{2\lambda_1 +
\lambda_2/2 + \lambda_3/2 - 2|\lambda_4|}
\end{equation}
The latter plays the role of the SM Higgs boson. 

The
remaining mass eigenstates ($\xi'_2$ and $\eta'_2$), which can
also be read off from Eq.  (\ref{neutral4}), are superpositions
of $\xi_2$ and $\eta_2$ defined through a rotation by the angle
$\alpha$ with masses 
\begin{equation}
m_{\xi_2', ~\eta_2'} = v \sqrt{|\lambda_4|} \;,\; \frac{v}{2}
\sqrt{3\lambda_2/2 - \lambda_3/2 + 2 |\lambda_4|} \;.
\label{eigen23}
\end{equation}

The other neutral scalars, namely $(\chi_3, \phi_3) \equiv
(\xi_3, \eta_3)$, are
decoupled from the rest and have the mass matrix:
\begin{equation}
M^2_{\chi_3, ~\phi_3} = \frac{v^2}{4} \left[
(-\frac{3}{4}\lambda_2 + \frac{1}{4}\lambda_3 + |\lambda_4|) I +
2|\lambda_4| \cos(3\alpha) \pmatrix{\cos\alpha & -\sin\alpha \cr
-\sin\alpha & -\cos\alpha  } \right] \;.
\label{neutral2a}
\end{equation}
The eigenvalues of this matrix are:
\begin{equation}
m_{\xi_3', ~\eta_3'} = \frac{v}{2} \sqrt{ -3\lambda_2/4
+ \lambda_3/4 + |\lambda_4| \{1 \mp
2\cos(3\alpha)\}} \;,
\label{eigen33}
\end{equation}
where $\xi_3'$ and $\eta_3'$ are obtained from $\xi_3$ and
$\eta_3$ by a rotation through an angle $\alpha/2$. 

A bound on the phase $\alpha$ 
is readily obtained from Eqs.
(\ref{eigen23}) and (\ref{eigen33}). One has:
\begin{equation}
m^2_{\xi_3', ~\eta_3'} = \frac{1}{2} m^2_{\xi_2'} \left[
\frac{m^2_{\xi_2'} - m^2_{\eta_2'}}{m^2_{\xi_2'}} \mp \cos(3\alpha)
\right] \;.
\end{equation} 
One can immediately conclude that:
\begin{equation}
|\cos(3\alpha)| \leq \frac{m^2_{\xi_2'} -
m^2_{\eta_2'}}{m^2_{\xi_2'}} \;.
\end{equation}
Thus one must have 
\begin{equation}
m^2_{\xi_2'} > m^2_{\eta_2'} \; \Rightarrow 2|\lambda_4| >
\frac{3}{2} \lambda_2 - \frac{1}{2} \lambda_3 \;. 
\end{equation}
In addition, one has the sum-rule:
\begin{equation}
m^2_{\eta'_2} + m^2_{\xi'_3} + m^2_{\eta'_3} =  m^2_{\xi'_2} \;,
\end{equation}

It is worth bearing in mind that the choices $\alpha = 0$ and
$\alpha = \pi$, which correspond to the real limit, are
inadmissible. In both cases $m_{\eta'_2}$ has to vanish, which is
ruled out on physics grounds.

It is not difficult to ensure the reality of all the scalar
masses at the same time by a suitable choice of the
$\lambda_i$ while satisfying the requirements of positivity of
the potential.



\end{document}